\pdfoutput=1

\documentclass[11pt]{article}
\usepackage{booktabs}
\usepackage{amssymb}
\usepackage[final]{acl}
\usepackage{amsmath}
\usepackage{times}
\usepackage{latexsym}
\usepackage{multirow}
\usepackage{multicol}
\usepackage{colortbl}
\usepackage{makecell}
\usepackage[T1]{fontenc}
\usepackage{subcaption}  
\usepackage[utf8]{inputenc}
\usepackage{fancyvrb} 
\usepackage{microtype}
\usepackage{multirow}
\usepackage{inconsolata}

\usepackage{graphicx}

\title{WavRAG: Audio-Integrated Retrieval Augmented Generation for Spoken Dialogue Models}

\author{
\hspace*{-0.6cm} 
Yifu Chen~$^{\spadesuit}$~\thanks{Equal contribution.}
~~Shengpeng Ji~$^{\spadesuit}$~\footnotemark[1]
~~Haoxiao Wang~$^{\spadesuit}$~\footnotemark[1]
~~Ziqing Wang~$^{\clubsuit}$
~~Siyu Chen ~$^{\spadesuit}$
\\
\textbf{
\hspace*{-0.4cm} 
Jinzheng He~$^{\heartsuit}$ 
~~Jin Xu~$^{\heartsuit}$ 
~~Zhou Zhao~$^\spadesuit$\thanks{Corresponding author.}
} \\
$^\spadesuit$~Zhejiang University~ $^{\heartsuit}$~Alibaba Group~
$^{\clubsuit}$ Beijing University of Technology\\
\texttt{Eve106298@163.com}\\ 
\texttt{zhaozhou@zju.edu.cn}
}

\begin{document}

\maketitle
\begin{abstract}
Retrieval Augmented Generation (RAG) has gained widespread adoption owing to its capacity to empower large language models (LLMs) to integrate external knowledge. However, existing RAG frameworks are primarily designed for text-based LLMs and rely on Automatic Speech Recognition to process speech input, which discards crucial audio information, risks transcription errors, and increases computational overhead. Therefore, we introduce WavRAG, the first retrieval augmented generation framework with native, end-to-end audio support. WavRAG offers two key features: 1) Bypassing ASR, WavRAG directly processes raw audio for both embedding and retrieval. 2) WavRAG integrates audio and text into a unified knowledge representation. Specifically, we propose the WavRetriever to facilitate the retrieval from a text-audio hybrid knowledge base, and further enhance the in-context capabilities of spoken dialogue models through the integration of chain-of-thought reasoning. In comparison to state-of-the-art ASR-Text RAG pipelines, WavRAG achieves comparable retrieval performance while delivering a 10x acceleration. Furthermore, WavRAG's unique text-audio hybrid retrieval capability extends the boundaries of RAG to the audio modality.
\end{abstract}

\section{Introduction}
\begin{figure*}[ht]
    \centering
    \includegraphics[width=\linewidth]{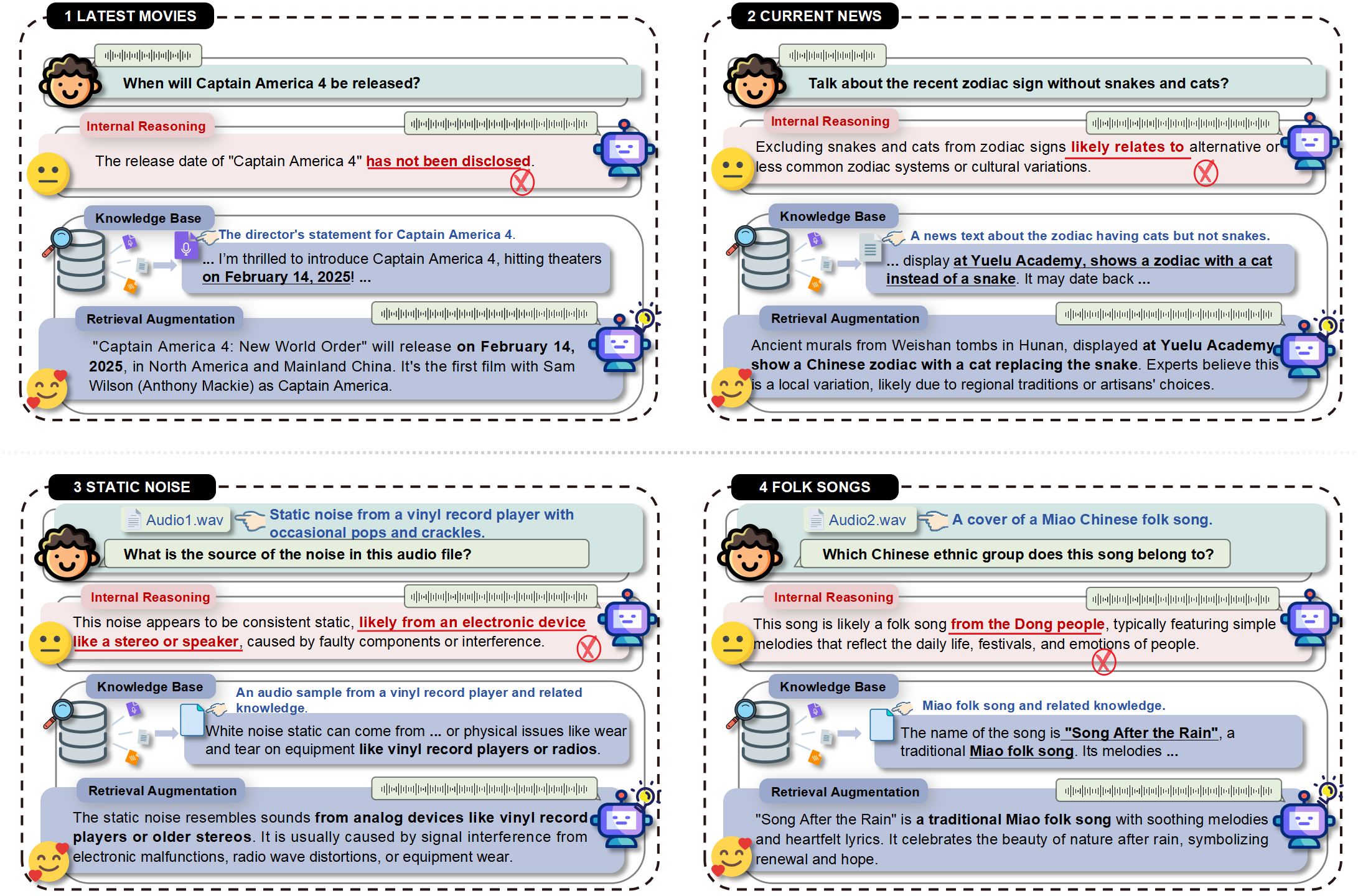}

    \caption{Dialogue examples illustrating WavRAG's ability to understand queries and generate appropriate responses by retrieving and augment relevant diverse modality  knowledge}
    \label{fig:img1}
\end{figure*}
Retrieval Augmented Generation (RAG)~\cite{guu2020realmretrievalaugmentedlanguagemodel} has revolutionized natural language processing, offering a powerful approach to enhance text-based large language models (LLMs).  A typical RAG system comprises three main components: a knowledge source (e.g., a database of documents or a knowledge graph)~\cite{peng2024graphretrievalaugmentedgenerationsurvey}, a retriever module, and a generator module (typically a large language model). The RAG process unfolds in two primary stages.
1) Retrieval: Given an input query, the retriever module identifies and retrieves relevant information from the knowledge source.
2) Generation: The retrieved information, along with the original input query, is provided as context to the generator, which then produces the final response. This process allows the LLM to leverage up-to-date and factual knowledge, significantly improving the accuracy, consistency, and relevance of its responses~\cite{gao2024retrievalaugmentedgenerationlargelanguage}. Researchers have sought to extend these benefits to the spoken dialogue domain~\cite{7114229,4490200}, aiming to enhance spoken dialogue models capable of processing audio inputs and generating speech responses~\cite{fang2024llamaomniseamlessspeechinteraction,ji2024wavchatsurveyspokendialogue}. However, these efforts have largely relied on cascaded "ASR+RAG" pipelines, which first transcribe speech to text using Automatic Speech Recognition (ASR) and then apply a text-based RAG system. This indirect approach suffers from several critical limitations: it fails to fully leverage the rich information present in the audio modality, treating it as a mere intermediary; the ASR component introduces computational overhead and potential transcription errors that propagate through the system; and the reliance on text-centric knowledge bases restricts the system's ability to utilize audio-specific knowledge. Crucially, the audio modality encompasses far more than just human speech; it includes a vast range of sounds~\cite{ji2024wavtokenizer}, such as environmental noises, music, and animal vocalizations, many of which are beyond the capabilities of ASR. A RAG framework that natively integrates this broader spectrum of audio information can unlock significant potential for richer, more contextually relevant understanding and generation~\cite{chen2022muragmultimodalretrievalaugmentedgenerator}, directly addressing key challenges of spoken dialogue models, such as the tendency to generate hallucinated content. However, realizing a fully end-to-end, audio-compatible RAG system remains a significant and open research challenge.

Addressing the limitations of existing approaches requires a fundamental shift: building knowledge bases that encompass a wide range of audio modalities alongside text, and developing retrieval mechanisms that can effectively represent and retrieve information from this unified audio-text space. Additionally,effectively harnessing retrieved multimodal information during generation requires new techniques for improved accuracy, naturalness, and contextual consistency. Therefore, we propose WavRAG, a novel end-to-end RAG framework designed for native audio integration. We show several dialogue scenarios to help understand the role of our framework in Figure~\ref{fig:img1}. Inspired by LLM2Vec's~\cite{behnamghader2024llm2veclargelanguagemodels} success in fine-tuning LLMs for text embeddings, we build our retriever on top of Qwen2-Audio, an MLLM with strong general audio comprehension, to create a unified embedding space for audio (speech and non-speech) and text. Considering that The pre-training objectives of multimodal language models are not optimized for retrieval, we further enhance the model with a contrastive learning framework. This approach allows the resulting retriever to encode end-to-end, directly encoding raw audio and text inputs into a shared embedding space, thus avoiding the computational overhead and potential error propagation of cascaded ASR-Text pipelines. Furthermore, in the generation stage, We incorporate Chain-of-Thought (CoT) reasoning, promoting a structured and interpretable inference process that enhances both reliability and controllability in utilizing retrieved multimodal knowledge.
In summary, our contributions are as follows:
\begin{itemize}

    \item We propose WavRAG, a novel RAG framework for spoken dialogue models. It is the first to extend RAG to this domain in an end-to-end manner and to incorporate a hybrid text-audio knowledge base.
\item We introduce a novel retriever WavRetriever, to support hybrid retrieval across text-audio modalities, and further enhance the in-context capabilities of the spoken dialogue models through Chain-of-Thought techniques.
\item WavRAG achieves comparable results to the SOTA text-based RAG models in text retrieval, while offering an average acceleration of 10 times. Moreover, hybrid text-audio retrieval provides WavRAG with new capabilities.

\end{itemize}

\section{Related Works}

\paragraph{Audio RAG.}
While Retrieval-Augmented Generation (RAG) has shown promise in audio-related tasks like captioning~\cite{koizumi2020audiocaptioningusingpretrained,zhao-etal-2023-generating}, text-to-audio generation~\cite{huang2023makeanaudiotexttoaudiogenerationpromptenhanced}, and music generation~\cite{gonzales-rudzicz-2024-retrieval}. However, while these efforts demonstrate the utility of retrieval in audio processing,  prior work primarily utilizes retrieval to enhance specific, isolated tasks with limited exploration of how retrieval-augmented techniques can benefit spoken dialogue models. Audio information itself carries rich semantic and acoustic imformation that can improve retrieval grounding, enhance response contextualization, and strengthen factual consistency. WavRAG, in contrast, integrates retrieval as a core component of a complete dialogue system. The combination of general audio support and end-to-end integration distinguishes WavRAG and represents a significant advancement towards truly audio-native, retrieval-augmented spoken dialogue systems.

\paragraph{Multimodal Retrieval.}
The increasing prevalence of multimedia applications and Retrieval-Augmented Generation (RAG) systems, fueled by Multimodal Language Models (MLLMs), has underscored the necessity for unified retrieval frameworks capable of managing diverse modalities. Traditional cross-modality retrieval methods often rely on pre-trained models such as CLAP~\cite{elizalde2022claplearningaudioconcepts} and CLIP, which use separate encoders for text and other modalities (e.g., UniVL-DR~\cite{liu2023universalvisionlanguagedenseretrieval} and UniIR~\cite{wei2023uniirtrainingbenchmarkinguniversal}). Other approaches enhance pre-trained text embeddings with audio encoders~\cite{min2025speechretrievalaugmentedgenerationautomatic}, but these often prioritize the semantic content of speech, overlooking important general audio. Such methods struggle to effectively capture the full spectrum of information from both speech and non-speech audio.
Recent advancements have highlighted the potential of Large Language Models (LLMs) and Supervised Fine-Tuning (SFT) for creating powerful, unified text representations~\cite{behnamghader2024llm2veclargelanguagemodels,lee2025nvembedimprovedtechniquestraining}. This methodology has been successfully extended to other modalities, with works like E5-V~\cite{jiang2024e5vuniversalembeddingsmultimodal} and VLM2VEC~\cite{jiang2025vlm2vectrainingvisionlanguagemodels} focusing on fine-tuning strategies for visual models. Furthermore, Zhang's research~\cite{zhang2024gmeimprovinguniversalmultimodal} demonstrates the feasibility of developing universal multimodal retrieval models using MLLMs. However, there has been limited exploration in the audio modality, prompting us to propose the first end-to-end audio-text multimodal retriever.

\begin{figure*}[htbp]
    \centering
    \includegraphics[width=\linewidth]{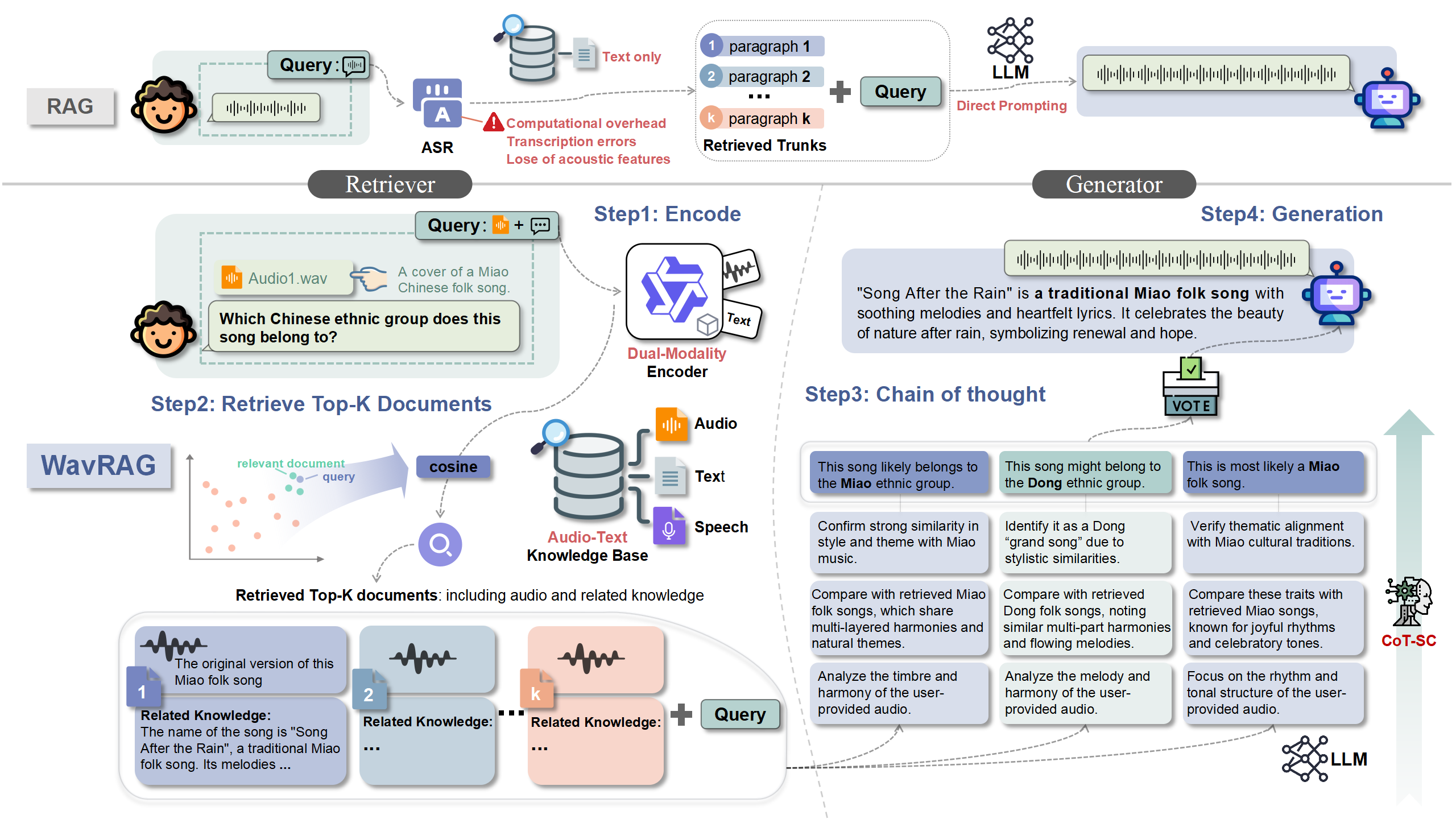}

    \caption{Architecture of the WavRAG framework. Top: Traditional RAG pipeline using ASR, highlighting its limitations. Bottom: WavRAG's four-step process: (1) A dual-modality encoder creates embeddings for both audio and text queries; (2) Top-K documents are retrieved from an audio-text knowledge base using cosine similarity; (3) A chain-of-thought reasoning process analyzes the retrieved information; (4) A large language model generates the final response, grounded in the retrieved knowledge.}
    \label{fig:img2}
\end{figure*}

\section{WavRAG}

\subsection{Overview}
Figure~\ref{fig:img2} provides an overview of the traditional text-based RAG framework (top) and our proposed WavRAG framework (bottom).

In the context of text-based dialogue models, the classic retrieval-augmented generation (RAG) framework, depicted in the top portion of Figure~\ref{fig:img2}, typically includes: (1) a text embedding model acting as the retriever  $R_{\boldsymbol{\phi}}$, (2) a text-based dialogue model~\cite{touvron2023llamaopenefficientfoundation} serving as the generator $G_{\theta}$, and (3) a fixed external knowledge corpus
$\mathcal{D} = \{ d_1, \ldots, d_N \}$
containing only textual snippets $d_i$.  During inference, the process is divided into two stages: retrieval and generation.

During retrieval, given a textual query $q_t$, the retriever computes the retrieval distribution
$p( d \mid q_i )$ via:

$$
p( d \mid q_i ) =\frac{\exp ( \text{sim}( R_{\phi}( q_i ) ,R_{\phi}( d ) ) )}{\sum_{d_i\in \mathcal{D}}{\exp ( \text{sim}( R_{\phi}( q_i ) ,R_{\phi}( d_i ) ) )}}
$$
where $R_{\phi}( \cdot ) $ denotes the encoding function of $R_{\boldsymbol{\phi}}$,
$\text{sim}( \cdot ,\cdot )$ is a similarity metric (such as cosine similarity, as shown in Figure~\ref{fig:img2}), and $d\in \mathcal{D}$. For efficiency,
$\mathcal{D}$ is usually pre-encoded offline by  $R_{\boldsymbol{\phi}}$. By drawing from this distribution, the system selects the Top-$k$ relevant snippets, forming a subset $D_k \subset \mathcal{D}$.

During generation, conditioned on $q_i$ and the retrieved subset $D_k$, the system computes the probability of producing the target text $y_i$ as follows:
$$
p( y_i \mid q_i,D_k ) =\prod_{m=1}^N{p( y_i \mid q_i,D_k,y_{<m} )}
$$
where $p( y_i \mid q_t,D_k,y_{<i} )$ is given by the generator $G_{\theta}$ and $N$ denotes the number of tokens in the answer $y_i$.  This process is illustrated in the top-right portion of Figure~\ref{fig:img2}, where the retrieved text snippets and the query are used to prompt the LLM generator.

To extend RAG to spoken dialogue scenes and overcome the limitations of the traditional ASR-based approach, we propose WavRAG, as shown in the bottom portion of Figure~\ref{fig:img2}. WavRAG integrates a retriever $R_{\boldsymbol{\phi}}$ capable of directly processing queries in audio, text, and combined formats, and interfacing with a multimodal knowledge corpus $K$. This design preserves the full information present in the audio, including both speech and non-speech sounds. Specifically, we extend the original textual query \(q_t\) and the knowledge corpus \(\mathcal{D}\) into a unified query \(q_{uni}\) and knowledge corpus \(K = \{ k_1, \ldots, k_i \}\), each of which may consist of audio, text, or a combination of these modalities, as depicted in the "Retriever" and "Knowledge Base" sections of Figure~\ref{fig:img2} (bottom).  At the generation stage (bottom-right of Figure~\ref{fig:img2}), we further introduce Chain-of-Thought (CoT) reasoning to systematically integrate external knowledge with the original input, ultimately producing the final answer.  We will elaborate on the retrieval and generation components in detail in the following sections.

\subsection{WavRetriever}
\begin{figure}[ht]
    \centering
    \includegraphics[width=\linewidth]{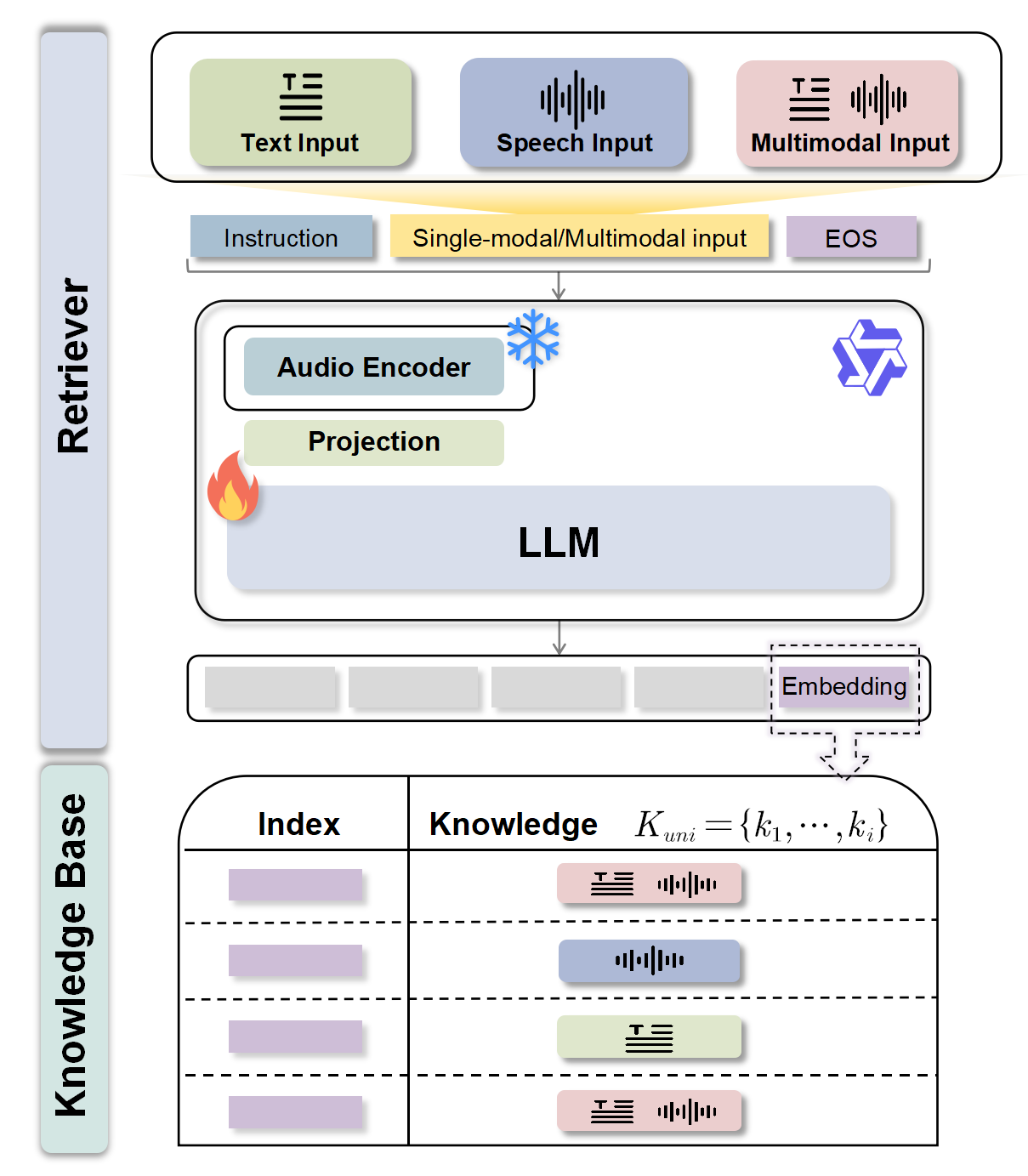}

    \caption{Architecture of the proposed multimodal retriever, showing the input processing, LLM-based encoding, and knowledge base structure.}
    \label{fig:img3-2}
\end{figure}

The goal of our retriever, \(R_{\boldsymbol{\phi}}\) (WavRetriever), is to produce embedding vectors for both queries and knowledge entries that enable efficient similarity-based retrieval. As depicted in Figure~\ref{fig:img3-2}, WavRetriever processes text, speech, or multimodal inputs, which are concatenated with a task-specific instruction and an End-of-Sequence (EOS) token. We build WavRetriever upon the Qwen2-Audio MLLM, leveraging its robust general audio comprehension. Specifically, we freeze the pre-trained audio encoder parameters of Qwen2-Audio and focus training on the projection layer and the backbone LLM. This allows us to capitalize on Qwen2-Audio's existing audio processing capabilities.

However, simply fine-tuning Qwen2-Audio on the downstream task is insufficient for optimal retrieval performance.  While pre-trained MLLMs like Qwen2-Audio possess robust multimodal understanding, their pre-training objectives are not directly optimized for creating embeddings suitable for similarity-based retrieval. To address this, we further adapt Qwen2-Audio into a powerful multimodal encoder using a carefully designed contrastive learning strategy. This strategy shapes the embedding space by maximizing the similarity between a query's embedding and its relevant (positive) knowledge, while minimizing the similarity with irrelevant (negative) knowledge embeddings.

Each training instance in our contrastive learning setup comprises a query, \(q_{ins} = \textit{Instruction: } \{\textit{prompt}\} \textit{ Query: } q_{uni}\), a positive knowledge sample \(k^+\), and a set of negative knowledge samples \(\{k^-_1, \ldots, k^-_l\}\).  Both the query, \(q_{uni}\), and the knowledge samples, \(k\), can be audio, text, or a combination thereof. We construct negative samples using in-batch negatives.  The representation for each instance is derived from the final hidden state of the last token. We employ the InfoNCE loss function~\cite{oord2019representationlearningcontrastivepredictive}, formulated as:

\begin{equation}
\label{eq:Z}
Z = \sum_{i=0}^{t} \exp \left( \frac{\text{sim}(r_q, r_{k,i})}{\tau} \right)
\end{equation}

\begin{equation}
\mathcal{L} = - \left[ \frac{\text{sim}(r_q, r_k^+)}{\tau} - \log Z \right]
\end{equation}

where \(sim(\cdot,\cdot)\) denotes cosine similarity, \(\tau\) is a temperature parameter, and \(r_q\) represents the embedding of the query obtained after processing by the retriever \(R_\phi\). In Equation~\eqref{eq:Z}, the index \(i=0\) represents the embedding of the positive knowledge sample \(r_k^+\), while the other \(i\) values correspond to the embeddings of the \(i\)-th negative knowledge sample \(r_{k,i}^-\). The multimodal knowledge entries are stored in the knowledge base, as shown at the bottom of Figure~\ref{fig:img3-2}.  Detailed training setup and dataset-specific prompts are provided in Appendix~\ref{sec:hyper}.

\begin{table*}[ht!]
\centering
\renewcommand{\arraystretch}{1.3} 

\resizebox{\textwidth}{!}{
\begin{tabular}{cccccccccc}
\Xhline{1.2pt}
\multirow[c]{2}{*}{\centering \arraybackslash \textbf{Task}} 
& \multirow[c]{2}{*}{\centering \arraybackslash \textbf{Dataset}} 
& \multirow[c]{2}{*}{\centering \arraybackslash \textbf{Model}} 
& \multirow[c]{2}{*}{\centering \arraybackslash \textbf{Whisper size}} 
& \multirow[c]{2}{*}{\centering \arraybackslash \textbf{WER}} 
& \multirow[c]{2}{*}{\centering \arraybackslash \textbf{Avg. Time}} 
& \multicolumn{4}{c}{\textbf{Metric}} \\ 
\cline{7-10}
& & & & & & \centering \arraybackslash \textbf{R@1} & \centering \arraybackslash \textbf{R@5} & \centering \arraybackslash \textbf{R@10} & \centering \arraybackslash \textbf{NDCG@10} \\
\Xhline{1.2pt}

\multirow[c]{5}{*}{\centering \arraybackslash \textbf{Speech2Text}} 
& \multirow[c]{5}{*}{\centering \arraybackslash HotpotQA}
& \multirow[c]{3}{*}{\centering \arraybackslash BGE} 
& Tiny   
  & 37.55\% & 1.26 
  & 0.3741 & 0.7024 & 0.7509 & 0.4628 \\
& & & Medium 
  & 21.67\% & 1.48
  & 0.4440 & 0.8319 & 0.8736 & 0.5190 \\
& & & Large  
  & 19.2\%  & 1.92
  & \textbf{0.4533} & \textbf{0.8519} & 0.8895 & \textbf{0.5252} \\
\rowcolor[HTML]{EFEFEF}
& & \textbf{Ours} 
  & - & - & \textbf{0.23}
  & 0.4532 & 0.8492 & \textbf{0.8898} & 0.5117 \\
\rowcolor[HTML]{FFF4E0}
\multicolumn{10}{c}{
\textbf{Comparison vs.\ BGE (Tiny / Medium / Large):} 
\quad \textit{Speed-up} $\approx 5.49\times / 6.43\times / 8.35\times,$ 
\quad $\Delta R@\!10 \approx +0.139 / +0.016 / +0.0003.$
} \\
\hline

\multirow[c]{3}{*}{\centering \arraybackslash \textbf{Text2Speech}} 
& \multirow[c]{3}{*}{\centering \arraybackslash Spoken-SQuAD}
& CLSR  
  & - & - & -
  & 0.4982 & 0.7963 & 0.8583 & - \\
& & BGE  
  & - & 44.22\% & -
  & 0.5464 & 0.7767 & 0.8497 & 0.6947 \\
\rowcolor[HTML]{EFEFEF}
& & \textbf{Ours} 
  & - & - & \textbf{0.11}
  & \textbf{0.6844} & \textbf{0.8374} & \textbf{0.9023} & \textbf{0.8483} \\
\hline

\multirow[c]{6}{*}{\centering \arraybackslash \textbf{Speech2Speech}}
& \multirow[c]{6}{*}{\centering \arraybackslash SLUE-SQA-5}
& CLSR 
  & - & 16.69\% &
  & 0.3065 & 0.6219 & \textbf{0.7443} & - \\
& & \multirow[c]{3}{*}{\centering \arraybackslash BGE}
& Tiny
  & 45.34\%/53.66\% & 0.62/1.27
  & 0.1696 & 0.3871 & 0.4828 & 0.2194 \\
& & & Medium
  & 26.14\%/44.46\% & 0.87/3.44
  & 0.3228 & 0.5940 & 0.6982 & 0.2989 \\
& & & Large
  & 23.59\%/42.19\% & 0.98/4.63
  & 0.3312 & 0.6121 & 0.7196 & 0.3269 \\
\rowcolor[HTML]{EFEFEF}
& & \textbf{Ours} 
  & - & - & \textbf{0.17/0.22}
  & \textbf{0.3392} & \textbf{0.6308} & 0.7221 & \textbf{0.3623} \\
\rowcolor[HTML]{FFF4E0}
\multicolumn{10}{c}{
\textbf{Comparison vs.\ BGE (Tiny / Medium / Large):}
\quad \textit{Speed-up} $\approx 4.84\times /11.05\times / 14.38\times,$ 
\quad $\Delta R@\!10 \approx +0.2393 / +0.0282 / +0.0025.$
} \\
\hline

\multirow[c]{6}{*}{\centering \arraybackslash \textbf{Audio+Text2Audio+Text}}
& \multirow[c]{6}{*}{\centering \arraybackslash Ours}
& ClAP (AT) 
  & - & - & \textbf{0.05}
  & 0.1260 & 0.2940 & 0.3989 & 0.2474 \\
& & CLAP(TA) 
  & - & - & \textbf{0.05}
  & 0.0998 & 0.2577 & 0.3588 & 0.2135 \\
& & CLAP(AT2AT) 
  & - & - & 0.09
  & 0.1345 & 0.2145 & 0.2379 & 0.1849 \\
& & ClAP (ALL) 
  & - & - & 0.06
  & 0.0001 & 0.0012 & 0.0018 & 0.0002 \\
& & BGE (Caption) 
  & - & - & 1.99
  & 0.0251 & 0.0585 & 0.0775 & 0.0483 \\
\rowcolor[HTML]{EFEFEF}
& & \textbf{Ours} 
  & - & - & 0.19
  & \textbf{0.2728} & \textbf{0.5184} & \textbf{0.6313} & \textbf{0.4381} \\
\Xhline{1.2pt}
\end{tabular}
} 

\caption{Comparison of various models and configurations across multiple tasks including Speech2Text, Text2Speech, Speech2Speech, and  Audio+Text2Audio+Text . Performance metrics include Word Error Rate (WER), Average Time, and multiple retrieval metrics (R@1, R@5, R@10, and nDCG@10). CLAP are evaluated for their respective tasks: CIAP (AT) is tested only on the Audio-to-Text retrieval task subset, CLAP (TA) on the Text-to-Audio retrieval task, CLAP (AT2AT) on Audio-to-Text-to-Audio tasks, and CLAP (ALL) on the entire dataset. Speed-up and performance changes relative to BGE (different Whisper model size Tiny/Medium/Large) configurations are reported.}
\label{tab:retrival_experiment}
\end{table*}

\subsection{Generation}

In WavRAG's generation stage, we adopt a retrieval-augmented generation paradigm.  The retriever provides the top-\(k\) retrieved knowledge entries (which can be audio, text, or multimodal) along with the original query \(q_{uni}\) as input to the generator.  While this approach provides rich contextual information, most existing spoken dialogue systems are not trained on these lengthy and multiple mixed-modality input formats, making naive concatenation of all retrieved documents prone to suboptimal performance. To address this, we incorporate Chain-of-Thought (CoT) reasoning, specifically Zero-Shot-CoT~\cite{kojima2022large} and a Self-Consistency mechanism.

\paragraph{Zero-Shot-CoT Reasoning.}
Zero-Shot-CoT prompting leverages the in-context reasoning abilities of large language models (LLMs) to generate intermediate reasoning steps without requiring task-specific training examples.  Given the multimodal query \(q_{uni}\), a guiding prompt  \(P_{\text{prompt}}\), a "magic prompt" \(P'\) (e.g., "Let's think step-by-step"), and the top-\(k\) retrieved knowledge snippets \(\mathcal{K}_k\), the generator \(G_{\text{reasoning}}\) produces a reasoning chain \(C_{\text{answer}}\):

\begin{equation}
    C_{\text{answer}} = G_{\text{reasoning}}(q_{\text{uni}}, P_{\text{prompt}} + P', \mathcal{K}_k)
\end{equation}

The retrieved knowledge \(\mathcal{K}_k\) provides the context for the reasoning process, allowing the model to generate a logical, step-by-step deduction leading to the final answer.

\paragraph{Self-Consistency.}

To further enhance the reliability of the reasoning process, we employ a Self-Consistency mechanism.  This approach samples multiple reasoning paths from the LLM and then selects the most consistent answer among them.  This mitigates the risk of errors that can arise from relying on a single, potentially suboptimal, reasoning path.
Specifically, we use the Universal Self-Consistency (USC) method~\cite{chen2023universal}.  Instead of simply taking a majority vote among the generated answers (which can be problematic for free-form answers), USC concatenates all sampled reasoning paths and answers and prompts the LLM *itself* to select the most consistent response.  This leverages the LLM's own understanding to determine the best answer, given the multiple reasoning paths.

\section{Experiments}

\subsection{Datasets}
For training, we curated a dataset of 1.5M samples across five retrieval scenarios:
 \textbf{Speech-to-Text:} We adapted existing text retrieval datasets (e.g. HotpotQA~\cite{yang2018hotpotqadatasetdiverseexplainable}, Quora~\cite{wang2017bilateralmultiperspectivematchingnatural}) by synthesizing speech queries using the CosyVoice2 TTS model~\cite{du2024cosyvoice2scalablestreaming,an2024funaudiollm} with diverse voice prompts and noise augmentation.
\textbf{Speech-to-Speech and Text-to-Speech:} We used existing datasets: SLUE-SQA-5~\cite{shon-etal-2023-slue} and Spoken-SQuAD~\cite{li2018spokensquadstudymitigating}.
\textbf{Text-to-Text:} We used  existing text retrieval datasets: ELI5~\cite{fan-etal-2019-eli5}, NQ~\cite{kwiatkowski-etal-2019-natural}, HotpotQA~\cite{yang2018hotpotqadatasetdiverseexplainable}, MS MARCO, Quora, SQuAD~\cite{rajpurkar2016squad100000questionsmachine}, and TriviaQA~\cite{joshi-etal-2017-triviaqa}.
\textbf{Audio+Text-to-Audio+Text:} We process new data from sources like AudioSetSL, AudioCaps~\cite{kim-etal-2019-audiocaps}, MusicCaps~\cite{agostinelli2023musiclmgeneratingmusictext}, Clotho~\cite{drossos2019clothoaudiocaptioningdataset}, VoxCeleb~\cite{Nagrani_2017}, and Xeno-canto, where queries and documents are both general audio-text pairs. For evaluation, We test WavRAG on four datasets: HotpotQA, Spoken-SQuAD, SLUE-SQA-5, and our custom mixed modality dataset. Detailed data processing procedures, dataset statistics, and examples are provided in Appendix~\ref{app:train_data}


\subsection{Baselines}

\paragraph{Retrieval Baselines.} 1)~\textbf{BGE}~\cite{li2024makingtextembeddersfewshot}: A state-of-the-art text embedding model, used within an ASR-based pipeline for speech-related  retrieval tasks. 2)~\textbf{CLSR}~\cite{anonymous2025clsr}: A state-of-the-art speech-text retrieval framework. Used for comparison on speech-to-text and text-tp-speech retrieval task.
3)~\textbf{CLAP}~\cite{elizalde2022claplearningaudioconcepts}:  Used for comparison on our custom multimodal dataset.
4)~\textbf{Qwen2Audio-enhanced Text Retrieval}:  Used on our custom dataset. This baseline leverages Qwen2Audio to generate descriptive text from audio clips, which is then concatenated with the original text by using the BGE model.

\paragraph{Generation Baselines.}
1)~\textbf{TextRAG}: A standard text-based RAG pipeline using BGE embeddings for retrieval and Whisper medium for ASR. 

\begin{table}[ht]
\renewcommand{\arraystretch}{1.15}
\footnotesize
\resizebox{\columnwidth}{!}{%
\begin{tabular}{llcccccc}
\Xhline{1.3pt}
\multirow{2}{*}{Method} & \multirow{2}{*}{Model} & \multirow{2}{*}{Input} & \multicolumn{2}{c}{EM} & \multirow{2}{*}{Avg EM} & \multirow{2}{*}{FS} \\
\cmidrule(lr){4-5}
                      &                        &                       & HotpotQA & SLUE-SQA-5 &                    & (Ours)  \\
\midrule
\multicolumn{7}{c}{\textbf{(a) TextRAG}} \\
\midrule
\multirow{4}{*}{GPT-4o} 
 &  & top-1   & 0.3124 & 0.3237 & 0.3181  & -      \\
 &  & top-2   & 0.3457 & 0.3359 & 0.3408  & -      \\
 &  & top-3   & 0.3623 & 0.3531 & 0.3577  & -      \\
 &  & \cellcolor{gray!40}{Oracle}  & 0.5853 & 0.5931 & 0.5892      & -      \\
\midrule
\multirow{4}{*}{QwenAudio} 
 &  & top-1   & 0.1783 & 0.2439 & 0.2111  & -      \\
 &  & top-2   & 0.2336 & 0.2502 & 0.2419  & -      \\
 &  & top-3   & 0.2417 & 0.2561 & 0.2489  & -      \\
 &  & \cellcolor{gray!40}{Oracle}  & 0.4867 & 0.4784 & 0.4824      & -      \\
\midrule
\multicolumn{7}{c}{\textbf{(b) WavRAG}} \\
\midrule
\multirow{4}{*}{GPT-4o} 
 &  & top-1   & 0.4019 & 0.3904 & \cellcolor{blue!10}\textbf{0.3962}  & \cellcolor{blue!10}\textbf{0.5732} \\
 &  & top-2   & 0.4186 & 0.4315 & \cellcolor{blue!10}\textbf{0.4249}  & \cellcolor{blue!10}\textbf{0.6408} \\
 &  & top-3   & 0.4271 & 0.4007 & \cellcolor{blue!10}\textbf{0.4139}  & \cellcolor{blue!10}\textbf{0.5129} \\
 &  & \cellcolor{gray!40}{Oracle}  & 0.5941 & 0.6164 & \cellcolor{blue!10}\textbf{0.6053}  & \cellcolor{blue!10}\textbf{0.7096} \\
\midrule
\multirow{4}{*}{QwenAudio} 
 &  & top-1   & 0.2033    & 0.2647     & \cellcolor{blue!10}\textbf{0.2340}      & \cellcolor{blue!10}\textbf{0.5387} \\
 &  & top-2   & 0.2439     & 0.2956     & \cellcolor{blue!10}\textbf{0.2698}      & \cellcolor{blue!10}\textbf{0.5521} \\
 &  & top-3   & 0.2658     & 0.3063     & \cellcolor{blue!10}\textbf{0.2860}     & \cellcolor{blue!10}\textbf{0.5387} \\
 &  & \cellcolor{gray!40}{Oracle}  & 0.5032     & 0.5294    & \cellcolor{blue!10}\textbf{0.5163}      & \cellcolor{blue!10}\textbf{0.6079} \\
\midrule
\multicolumn{7}{c}{\textbf{(c) WavRAG-CoT}} \\
\midrule
\multirow{4}{*}{GPT-4o} 
 &  & top-1   & 0.4261 & 0.4520 & \cellcolor{blue!10}\textbf{0.4390}  & \cellcolor{blue!10}\textbf{0.6412} \\
 &  & top-2   & 0.4286 & 0.5239 & \cellcolor{blue!10}\textbf{0.4983}  & \cellcolor{blue!10}\textbf{0.6487} \\
 &  & top-3   & 0.4403 & 0.4918 & \cellcolor{blue!10}\textbf{0.4662}  & \cellcolor{blue!10}\textbf{0.5981} \\
 &  & \cellcolor{gray!40}{Oracle}  & 0.5976 & 0.6849 & \cellcolor{blue!10}\textbf{0.6413}  & \cellcolor{blue!10}\textbf{0.7389} \\
\midrule
\multirow{4}{*}{QwenAudio} 
 &  & top-1   & 0.2688     & 0.3132      & \cellcolor{blue!10}\textbf{0.2910}     & \cellcolor{blue!10}\textbf{0.6386} \\
 &  & top-2   & 0.3026     & 0.3352     & \cellcolor{blue!10}\textbf{0.3189}      & \cellcolor{blue!10}\textbf{0.6017} \\
 &  & top-3   & 0.3152    & 0.3397     & \cellcolor{blue!10}\textbf{0.3275}      & \cellcolor{blue!10}\textbf{0.5612} \\
 &  & \cellcolor{gray!40}{Oracle}  & 0.5863     & 0.6103     & \cellcolor{blue!10}\textbf{0.5983}      & \cellcolor{blue!10}\textbf{0.7122} \\
\Xhline{1.3pt}
\end{tabular}%
}
\caption{Generation experiment results: Performance comparison of TextRAG, WavRAG, and WavRAG-CoT on HotpotQA and SLUE-SQA-5 datasets. Metrics include Exact Match (EM) and F1-Score (FS). Results are shown for both GPT-4o and QwenAudio base models, with varying numbers of retrieved documents (top-1, top-2, top-3, and Oracle).}
\label{tab:generation_performance}
\end{table}

\subsection{Evaluation Metrics and Experimental Settings}
\paragraph{Retrieval.}
Retrieval performance is evaluated across four scenarios: Speech-to-Text (HotpotQA), Speech-to-Speech (SLUE-SQA-5), Text-to-Speech (Spoken-SQuAD), and Audio+Text to Audio+Text (a custom dataset). Reported metrics include:

Recall@k: The proportion of relevant items found within the top-k retrieved results.  Higher is better.  (We report Recall@1, Recall@5, and Recall@10).
NDCG@10 (Normalized Discounted Cumulative Gain): A measure of ranking quality that considers the position of relevant items in the retrieved list, giving higher scores to relevant items ranked higher. Higher is better.
Average Inference Time: The average time taken to process a single query.

For scenarios involving speech input, the Word Error Rate (WER) and model size of the Whisper ASR model (used in baseline methods) are also reported.  WER measures the accuracy of the speech recognition, with lower values indicating fewer errors.

\paragraph{Generation.}
Three RAG frameworks are compared: TextRAG (which uses Whisper medium for ASR and BGE embeddings for retrieval), WavRAG (using WavRetriever), and WavRAG-CoT (WavRAG with Chain-of-Thought reasoning). Both GPT-4o and QwenAudio serve as the generation models. The following metrics are used:

EM (Exact Match): For short-form answers (in HotpotQA and SLUE-SQA-5), this is a binary metric. It is 1 if the generated answer exactly matches the ground truth answer and 0 otherwise. Higher is better.
FactScore: For long-form answers (in the custom dataset), this metric, utilizing Qwen-plus for fact verification, evaluates the factual accuracy of the generated text by assessing the proportion of factual claims in the generation that are supported by the retrieved evidence. Higher is better.

Results are presented for the top-1, top-2, and top-3 retrieved documents. This means the generation model is provided with the single most relevant retrieved document (top-1), the two most relevant documents (top-2), and the three most relevant documents (top-3), respectively. Results are also presented for an "Oracle" condition where only the ground truth document is used.
\begin{table}[ht]
\centering
\resizebox{\columnwidth}{!}{\begin{tabular}{llccc}
\toprule
\textbf{Dataset} & \textbf{Metric} & \textbf{Qwen2audio (Original)} & \textbf{WavRAG} & \textbf{Improvement} \\
\midrule
\multirow{4}{*}{Ours} 
 & R@1     & 0.0675  & \textbf{0.2728} & \textcolor{blue}{\textit{+0.2053}} \\
 & R@5     & 0.1457  & \textbf{0.5184} & \textcolor{blue}{\textit{+0.3727}} \\
 & R@10    & 0.1868  & \textbf{0.6313} & \textcolor{blue}{\textit{+0.4445}} \\
 & nDCG@10 & 0.1212  & \textbf{0.5381} & \textcolor{blue}{\textit{+0.4169}} \\
\midrule
\multirow{4}{*}{Spoken-SQuAD} 
 & R@1     & 0.3407  & \textbf{0.6844} & \textcolor{blue}{\textit{+0.3437}} \\
 & R@5     & 0.4995  & \textbf{0.8374} & \textcolor{blue}{\textit{+0.3379}} \\
 & R@10    & 0.6003  & \textbf{0.9023} & \textcolor{blue}{\textit{+0.3020}} \\
 & nDCG@10 & 0.3554  & \textbf{0.8483} & \textcolor{blue}{\textit{+0.4929}} \\
\midrule
\multirow{4}{*}{HotpotQA} 
 & R@1     & 0.1457  & \textbf{0.4532} & \textcolor{blue}{\textit{+0.3075}} \\
 & R@5     & 0.3172  & \textbf{0.8492} & \textcolor{blue}{\textit{+0.5320}} \\
 & R@10    & 0.3858  & \textbf{0.8898} & \textcolor{blue}{\textit{+0.5040}} \\
 & nDCG@10 & 0.2868  & \textbf{0.5117} & \textcolor{blue}{\textit{+0.2249}} \\
\bottomrule
\end{tabular}}
\caption{Performance improvement after contrastive learing }
\label{tab:ablation}
\end{table}
\begin{figure}[ht]
    \centering
    \includegraphics[width=\linewidth]{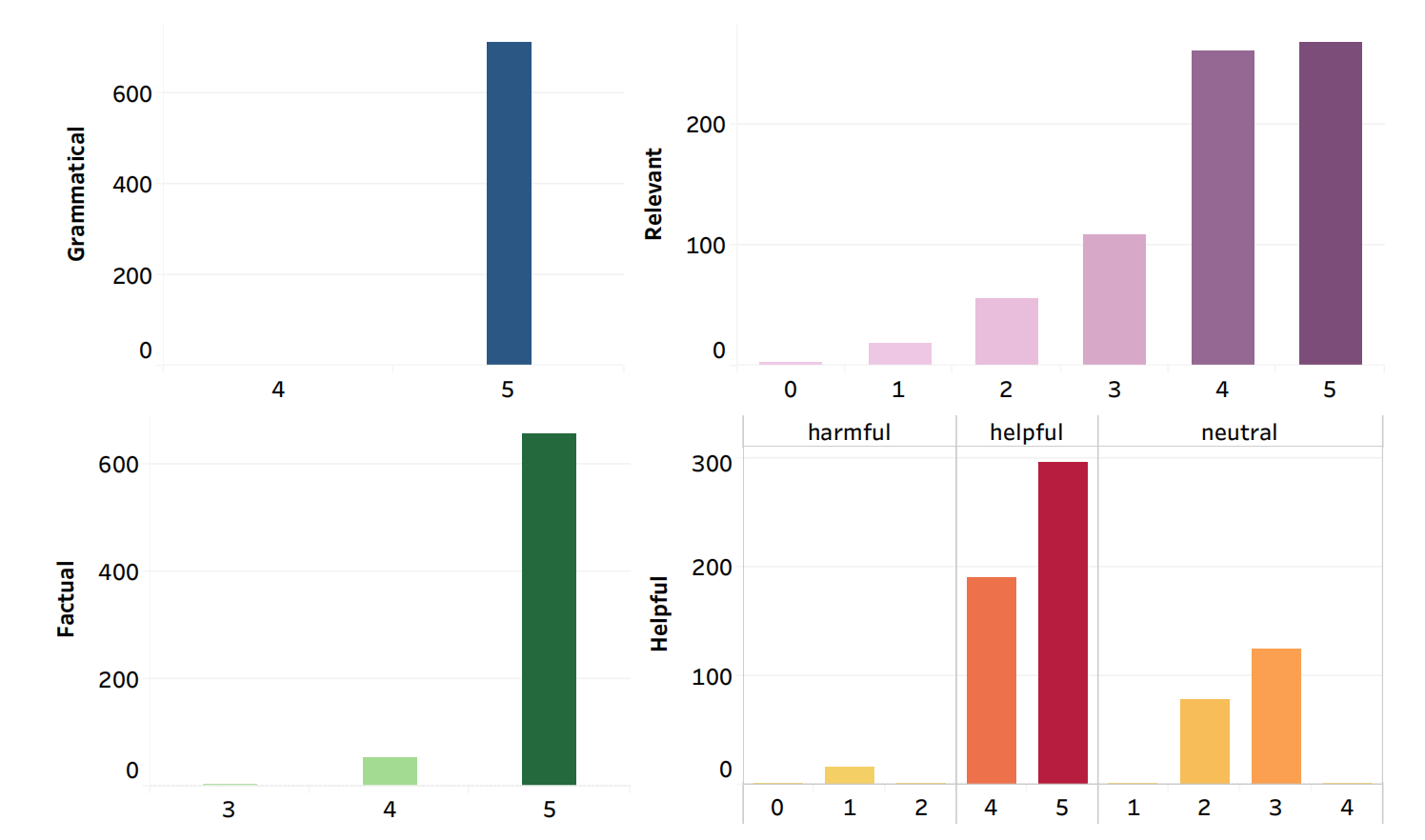}

    \caption{Human evaluation of knowledge quality. Distributions are shown for Grammatical scores, Factual scores, Relevance scores, and Helpfulness scores. The Helpfulness plot is further broken down by helpfulness level (helpful, neutral, harmful).}
    \label{fig:kb}
\end{figure}

\subsection{Main Results}

\paragraph{Retrieval Performance.}
Table~\ref{tab:retrival_experiment} presents the retrieval results, demonstrating WavRAG's superior performance and efficiency compared to traditional ASR-dependent baselines and other approaches across all four evaluated tasks: Speech-to-Text, Text-to-Speech, Speech-to-Speech, and Audio+Text-to-Audio+Text.

WavRAG's key advantage lies in its direct processing of audio inputs, eliminating the need for ASR and its associated computational overhead and potential for transcription errors. This translates to significant speedups in inference time, ranging from approximately 5x to over 14x compared to BGE models using Whisper, while simultaneously achieving comparable or superior retrieval accuracy. The most substantial performance gains are seen in the challenging Audio+Text-to-Audio+Text scenario, where WavRAG dramatically outperforms all baselines. This highlights the effectiveness of WavRAG's unified multimodal embedding space in capturing the complex relationships between audio and text, a capability lacking in approaches that rely on separate encoders or text-based representations. Even in scenarios where strong text-based baselines exist (e.g., Speech-to-Text), WavRAG achieves competitive results without the ASR bottleneck.

\paragraph{Generation Performance.}
Table~\ref{tab:generation_performance} presents the generation experiment results. The results clearly demonstrate two key findings: 1) Impact of Direct Audio Input (WavRAG vs. TextRAG):
Across all datasets and both LLMs, WavRAG consistently outperforms the traditional TextRAG approach. This highlights the benefits of providing the generator with direct access to the original audio modality, rather than relying on potentially lossy ASR transcriptions. For example, with GPT-4o on HotpotQA, WavRAG achieves a top-1 EM of 0.4019, compared to TextRAG's 0.3124 (an absolute improvement of +0.0895). Similar gains are observed on SLUE-SQA-5 and with QwenAudio. On our custom dataset, where FactScore is used, WavRAG (with GPT-4o) achieves a top-1 score of 0.5732, demonstrating a substantial availability for the inclusion of audio information. 2) Effectiveness of Chain-of-Thought (WavRAG-CoT):
The addition of Chain-of-Thought (CoT) reasoning further enhances performance. WavRAG-CoT consistently outperforms standard WavRAG across all datasets and models. With GPT-4o, the top-1 EM increases from 0.3904 to 0.4520 on SLUE-SQA-5 (+0.0616). On our custom dataset, the FactScore improves from 0.5732 to 0.6412 (+0.068). A noteworthy observation is the performance trend across Top-k configurations. We observe a decrease in performance when moving from Top-2 to Top-3 in our dataset.This degradation suggests Models struggle to prioritize and synthesize information from a larger, more heterogeneous set of inputs.The fact that WavRAG-CoT mitigates this issue is significant. The structured, step-by-step reasoning enforced by CoT provide a mechanism for the model to better manage the complexity of multimodal knowledge.

\subsection{Analysis}

\paragraph{Ablation Studies on Contrasitive Training Framework.}
To isolate the impact of our contrastive learning framework, we conducted an ablation study comparing the fine-tuned WavRAG retriever to the pre-fine-tuned Qwen2-Audio-7B-Instruct model. This baseline represents a strong, pre-trained MLLM with inherent multimodal understanding, but without retrieval-specific optimization. Following \cite{jiang2023scalingsentenceembeddingslarge}, original Qwen2-Audio's representation was obtained by prompting for global semantics at the last token.

Table~\ref{tab:ablation} shows the results, with WavRAG significantly outperforming the baseline across all datasets and metrics. Recall@1 improvements range from +0.3075 to +0.3437, and nDCG@10 gains are even more pronounced, reaching up to +0.4169. These substantial improvements unequivocally validate the effectiveness of our contrastive learning framework in adapting the MLLM for multimodal retrieval.
\paragraph{Knowledge Extension Quality.}
A critical aspect of WavRAG is its ability to take advantage of extended knowledge associated with audio. To assess the quality of this generated knowledge, we conducted a human evaluation on 700 randomly sampled instances from our custom multimodal dataset. 
Fluent English-speaking annotators evaluated each sample on a 5-point scale across four key dimensions: Grammaticality , Factual accuracy, Relevance , and overall Helpfulness (further categorized as helpful, neutral, or harmful).
Figure~\ref{fig:kb} shows the score distributions. The vast majority of samples received a score of 5 for Grammaticality, Factual accuracy, and Relevance, indicating high-quality knowledge generation. Furthermore, most samples were rated as "Helpful," demonstrating the positive impact of the extended knowledge on audio understanding.
 
\section{Conclusions}
This work introduced WavRAG, a novel retrieval-augmented generation framework specifically designed for spoken dialogue systems. WavRAG makes a significant departure from traditional ASR-dependent pipelines by directly processing raw audio input for  embedding and retrieval. This approach offers several key advantages, including reduced computational overhead, preservation of rich acoustic information, and the ability to leverage a unified multimodal knowledge base.Through comprehensive experiments, including quantitative evaluations and qualitative analyses , we demonstrated the effectiveness of WavRAG. The results show substantial improvements in both retrieval and generation performance compared to traditional methods and baseline models. 

\section*{Limitations}
Despite WavRAG's exploration of how a well-designed RAG system can leverage both semantic and acoustic information to enhance the semantic quality of responses, emotional tone and prosody are equally crucial in spoken dialogue systems. The extent to which RAG can contribute to the acoustic aspects of responses, such as intonation, expressiveness, and speaker style, remains an open question, warranting further investigation.

\bibliography{custom}

\appendix

\section{Training Detail and Task-specific Instrutions}
\label{sec:hyper}
In our training regimen, we employ Low-Rank Adaptation (LoRA)~\cite{hu2021loralowrankadaptationlarge} with a rank of 8, a learning rate of 1e-4, an alpha of 32, and a dropout rate of 0.05. To efficiently handle audio segments of varying lengths while maintaining training efficacy, each audio sample is tokenized with a maximum of 2000 tokens, and the corresponding text is limited to 512 tokens, with a batch size of 64. Additionally, gradient checkpointing is utilized and training is conducted in bfloat16 precision to optimize GPU memory usage; all experiments are executed on four NVIDIA A800 GPUs, each equipped with 80GB of memory. To further enhance model robustness to diverse acoustic environments, we integrate several speech data augmentation techniques that sequentially apply echo simulation, MUSAN noise addition, and random gain adjustments. Specifically, the echo effect is simulated by adding a delayed (100–500 ms) and scaled (0–0.2) version of the audio signal to itself; MUSAN noise is incorporated by randomly selecting a noise file, resampling it to the target rate if necessary, concatenating it to match the audio length, and mixing it at a target signal-to-noise ratio between -4 dB and 14 dB with a probability of 0.5; finally, random gain is applied with a probability of 0.5, scaling the audio by a factor corresponding to a gain between -4 dB and 15 dB.

\section{Data process pipeline and examples}
\label{app:train_data}
\subsection{Date process}
\begin{table}[]
    \renewcommand{\arraystretch}{1.5}

\resizebox{1\columnwidth}{!}{
\begin{tabular}{cclccc}
\Xhline{1.3pt}
\textbf{Task}                                                                                       & \multicolumn{2}{c}{\textbf{Dataset}} & \textbf{\begin{tabular}[c]{@{}c@{}}Train\\ Q-D pairs\end{tabular}} & \textbf{Retrieval Test} & \textbf{Generation Test} \\ \hline
\multirow{2}{*}{\textbf{Speech-to-Text}}                                                            & \multicolumn{2}{c}{Quora}            & 60202                                                              & -                       & -                        \\
                                                                                                    & \multicolumn{2}{c}{HotpotQA}         & 84516                                                              & 7405                    & 7405                     \\ \hline
\multirow{4}{*}{\textbf{Text-to-Text}}                                                              & \multicolumn{2}{c}{ELI5}             & 325475                                                             & -                       & -                        \\
                                                                                                    & \multicolumn{2}{c}{TrivialQA}        & 60315                                                              & -                       & -                        \\
                                                                                                    & \multicolumn{2}{c}{SQuAD}            & 87599                                                              & -                       & -                        \\
                                                                                                    & \multicolumn{2}{c}{MS MARCO}         & 485823                                                             & -                       & -                        \\ \hline
\textbf{Speech-to-Specch}                                                                           & \multicolumn{2}{c}{SLUE-SQA-5}       & 46186                                                              & 2382                    & 2382                     \\ \hline
\textbf{Text-to-Speech}                                                                             & \multicolumn{2}{c}{SpokenSQuAD}      & 37111                                                              & 5351                    & -                        \\ \hline
\multirow{8}{*}{\textbf{\begin{tabular}[c]{@{}c@{}}Audio+Text-to-Audio+Text\\ (Ours)\end{tabular}}} & \multicolumn{2}{c}{Ours}             & 78746                                                              & 8834                    & 1200                     \\
                                                                                                    & \multicolumn{2}{c}{AudioCaps}        & 35327                                                              & 4043                    & 572                      \\
                                                                                                    & \multicolumn{2}{c}{MusicCap}         & 4080                                                               & 442                     & 76                       \\
                                                                                                    & \multicolumn{2}{c}{Clotho}           & 2852                                                               & 314                     & 68                       \\
                                                                                                    & \multicolumn{2}{c}{VoxCeleb}         & 1091                                                               & 120                     & -                        \\
                                                                                                    & \multicolumn{2}{c}{Xeno-canto}       & 8771                                                               & 956                     & -                        \\
                                                                                                    & \multicolumn{2}{c}{Collected}        & -                                                                  & -                       & 43                       \\ \cline{2-6} 
                                                                                                    & \multicolumn{2}{c}{Total}            & 130867                                                             & 14709                   & 1959                     \\ \Xhline{1.3pt}
\end{tabular}
}
\caption{Training data}
\label{tab:train_data}
\end{table}

To address the challenges of audio-text retrieval across varied applications, we trained our model on four distinct retrieval scenarios: speech-to-text, speech-to-speech, text-to-speech and audio+text-to-audio+text,totally 1.5M samples, as shown in table~\ref{tab:train_data}. For each scenario, we meticulously designed a dedicated data construction pipeline, ensuring the training data appropriately reflects the task-specific nuances.

\paragraph{Speech-to-Text Retrieval}

For the speech-to-text retrieval task, we leveraged a suite of established text retrieval datasets, including ELI5~\cite{fan-etal-2019-eli5}, NQ~\cite{kwiatkowski-etal-2019-natural}, HotpotQA~\cite{yang2018hotpotqadatasetdiverseexplainable}, MS MARCO, Quora~\cite{wang2017bilateralmultiperspectivematchingnatural}, SQuAD~\cite{rajpurkar2016squad100000questionsmachine}, and TriviaQA~\cite{joshi-etal-2017-triviaqa}. To transform these text-based queries into speech, we employed the CosyVoice2~\cite{du2024cosyvoice2scalablestreaming} text-to-speech model~\cite{ji2024textrolspeech,ji2024controlspeech}. To further enrich the diversity and robustness of the synthesized speech, we incorporated several data augmentation techniques. Specifically, we randomly sampled voice prompts from the Common Voice 12.0 training dataset~\cite{ardila-etal-2020-common} to prompt the CosyVoice2 synthesis process, thereby introducing speaker variability. Post-synthesis, we applied noise injection, random gain adjustments, and echo augmentation to simulate real-world acoustic conditions and enhance the model's adaptability to noisy environments.

\paragraph{Text-to-Text Retrieval}

\paragraph{Speech-to-Speech and Text-to-Speech Retrieval}

In the speech-to-speech retrieval and Text-to-Speech scenario, our objective was to enable the model to directly learn matching relationships at the speech level. To this end, we adopted the SLUE-SQA-5~\cite{shon-etal-2023-slue} and Spoken-SQuAD datasets~\cite{li2018spokensquadstudymitigating}. 

\paragraph{Audio+Text-to-Audio+Text Retrieval}

For the more complex audio+text-to-audio+text retrieval task, which demands the integration of acoustic and semantic information, we curated and utilized six datasets: AudioSetSL, AudioCaps~\cite{kim-etal-2019-audiocaps},MusicCaps~\cite{agostinelli2023musiclmgeneratingmusictext}, Clotho~\cite{drossos2019clothoaudiocaptioningdataset}, VoxCeleb~\cite{Nagrani_2017}, and Xeno-canto (referred to as Xeno).  In processing MusicCaps, AudioSetSL, and Clotho, we first computed audio similarity scores using the CLAP model. We then selected audio pairs which have the highest similarity to each other as positive matches. Subsequently,we feed these paired audio samples—along with their corresponding textual descriptions and predefined prompts—into the Gemini1.5 Pro model to generate extended knowledge, questions, and answers.  Within each pair, one audio clip was designated as the "query audio," and the other as the "knowledge audio." The "knowledge audio" and its associated generated expanded knowledge served as contextual cues to guide the retrieval process for the "query audio" and the generated textual question. Consequently, the model is trained to retrieve the corresponding "knowledge audio" and expanded knowledge given a "query audio" and a textual question. For unpaired audio instances within the original datasets, we retained them for auxiliary tasks such as audio-to-text or text-to-audio retrieval, ensuring comprehensive data utilization. Furthermore, we implemented supplementary processing for the Xeno (bird species dataset) and VoxCeleb (celebrity voices dataset) datasets. In Xeno, we randomly selected samples per species and augmented them with species descriptions retrieved via the Google API. In VoxCeleb, we constructed paired samples for celebrity voices, supplementing them with biographical information obtained through Google API searches. We also proactively collected knowledge-intensive audio question-answering data, which often involves content with rich background knowledge, such as film production company intro music or diverse cover versions of famous songs. By associating these audio clips with relevant background information, we created audio-text pairs to further enhance the model's capacity to integrate complex acoustic scenes with external knowledge.
\label{data_display}
\subsection{Prompt for Knowledge Extension}
In this section,We show the entire prompt for our dataset's knowledge extension,question generation and answer in Figure~\ref{fig:prompt}.

\begin{figure*}[h]
\centering

\begin{minipage}{1.1\textwidth}
\begin{Verbatim}[fontsize=\small]
I need to build a scenario like this:
The user inputs an audio B and asks a question about its information
(such as paralinguistic information such as style, emotion or
characteristics, or the knowledge it represents). To assist in the
analysis, the system finds a similar audio A and its text description
in the knowledge base through audio similarity retrieval to assist
in answering.

I will give you two audio clips and their captions, You need to
complete the following tasks:

Knowledge Expansion
Based on the caption of Audio A, expand on professional and detailed
background information about the audio itself as well as its broader
context. The expansion should be factual and detailed — information
that a typical LLM may not readily have or answer with certainty —
and Knowledge must directly help answer the generated questions.
Write this expanded knowledge in a style similar to Wikipedia, ensuring
logical consistency. Integrate mentions of Audio B question's answer
naturally rather than inserting them forcefully. But it should not
explicitly mention the existence of audio B. Don’t add too much
irrelevant information.

Question Generation
Generate a question about audio B, reflecting the need for audio
analysis. Cannot mention any information in Audio B in the question and
the questions should ensure that the LLM cannot accurately answer them
based on its own knowledge. The generated questions must be answerable
according to the knowledge of audio A above. Audio B and the question
should form a retrieval pair that uniquely retrieves audio A and its
knowledge. You need to ensure that the question reflects the advantages
of audio plus text dual-modal retrieval. Text alone cannot accurately retrieve.

Answer Generation
It must be an answer that can be answered based on the knowledge provided
by audio A and the knowledge that a typical large model may have. Do not
disclose the existence of Audio A to the user. When forming an answer,
use knowledge from Audio A in conjunction with the features of Audio B.
Ensure the final answer aligns with facts and audio B's caption but should
not explicitly mention the existence of audio B's caption.

You only need to output content in the following format:
Rewrite knowledge:

Question generation:

Answer generation:
\end{Verbatim}
\end{minipage}
\caption{Prompt for extension}
\label{fig:prompt}
\end{figure*}
\begin{table*}[]
\resizebox{\linewidth}{!}{\begin{tabular}{ccccll}
\Xhline{1.3pt}
\textbf{Task}                                      & \multicolumn{2}{c}{\textbf{Dataset}} & \multicolumn{3}{c}{\textbf{Instuction}}                                                                                                                                                                                                                                                                            \\ \hline
\multirow{2}{*}{\textbf{Speech-to-Text}}           & \multicolumn{2}{c}{Quora}            & \multicolumn{3}{c}{Given a question, retrieve questions that are semantically equivalent to the given question}                                                                                                                                                                                                    \\
                                                   & \multicolumn{2}{c}{HotpotQA}         & \multicolumn{3}{c}{Given a multi-hop question, retrieve documents that can help answer the question.}                                                                                                                                                                                                              \\ \hline
\multirow{4}{*}{\textbf{Text-to-Text}}             & \multicolumn{2}{c}{ELI5}             & \multicolumn{3}{c}{Given a question, retrieve relevant documents that best answer the question.}                                                                                                                                                                                                                   \\
                                                   & \multicolumn{2}{c}{TrivialQA}        & \multicolumn{3}{c}{Given a question, retrieve relevant documents that best answer the question.}                                                                                                                                                                                                                   \\
                                                   & \multicolumn{2}{c}{SQuAD}            & \multicolumn{3}{c}{Given a question, retrieve relevant documents that best answer the question.}                                                                                                                                                                                                                   \\
                                                   & \multicolumn{2}{c}{MS MARCO}         & \multicolumn{3}{c}{Given a web search query, retrieve relevant passages that answer the query.}                                                                                                                                                                                                                    \\ \hline
\textbf{Speech-to-Specch}                          & \multicolumn{2}{c}{SLUE-SQA-5}       & \multicolumn{3}{c}{Please retrieve the most relevant speech in the document based on the following questions}                                                                                                                                                                                                      \\ \hline
\textbf{Text-to-Speech}                            & \multicolumn{2}{c}{SpokenSQuAD}      & \multicolumn{3}{c}{Based on the following text query, retrieve the most relevant speech.}                                                                                                                                                                                                                          \\ \hline
\multirow{7}{*}{\textbf{Audio+Text-to-Audio+Text}} & \multirow{7}{*}{Ours}  & AudioSetSL  & \multicolumn{3}{c}{\begin{tabular}[c]{@{}c@{}}Audio2Text:Based on the following audio, extract the most relevant text description\\ Text2Audio:Based on the following text description, extract the most relevant audio\\ AT2AT:Extract the most relevant knowledge based on the following questions\end{tabular}} \\
                                                   &                        & AudioCaps   & \multicolumn{3}{c}{\begin{tabular}[c]{@{}c@{}}Audio2Text:Based on the following audio, extract the most relevant text description\\ Text2Audio:Based on the following text description, extract the most relevant audio\\ AT2AT:Extract the most relevant knowledge based on the following questions\end{tabular}} \\
                                                   &                        & MusicCap    & \multicolumn{3}{c}{\begin{tabular}[c]{@{}c@{}}Audio2Text:Based on the following audio, extract the most relevant text description\\ Text2Audio:Based on the following text description, extract the most relevant audio\\ AT2AT:Extract the most relevant knowledge based on the following questions\end{tabular}} \\
                                                   &                        & Clotho      & \multicolumn{3}{c}{\begin{tabular}[c]{@{}c@{}}Audio2Text:Based on the following audio, extract the most relevant text description\\ Text2Audio:Based on the following text description, extract the most relevant audio\\ AT2AT:Extract the most relevant knowledge based on the following questions\end{tabular}} \\
                                                   &                        & VoxCeleb    & \multicolumn{3}{c}{Extract the most relevant knowledge based on the following questions}                                                                                                                                                                                                                           \\
                                                   &                        & Xeno-canto  & \multicolumn{3}{c}{Extract the most relevant knowledge based on the following questions}                                                                                                                                                                                                                           \\
                                                   &                        & Collected   & \multicolumn{3}{c}{Extract the most relevant knowledge based on the following questions}                                                                                                                                                                                                                           \\ \Xhline{1.3pt}
\end{tabular}}
\caption{Dataset-specific prompt }
\end{table*}

\subsection{Data Samples Display}
This section presents several data samples from different datasets (see Tables~\ref{musicap1}, \ref{musiccap2}, \ref{collected}, \ref{vox}, and \ref{clotho}). Each sample is displayed in its own table formatted with two columns—one for the Field and one for the Content.

\subsection{Representation Visualization}
\begin{figure*}[ht]
    \centering
    \begin{subfigure}[b]{0.32\textwidth}
         \centering
         \includegraphics[width=\textwidth]{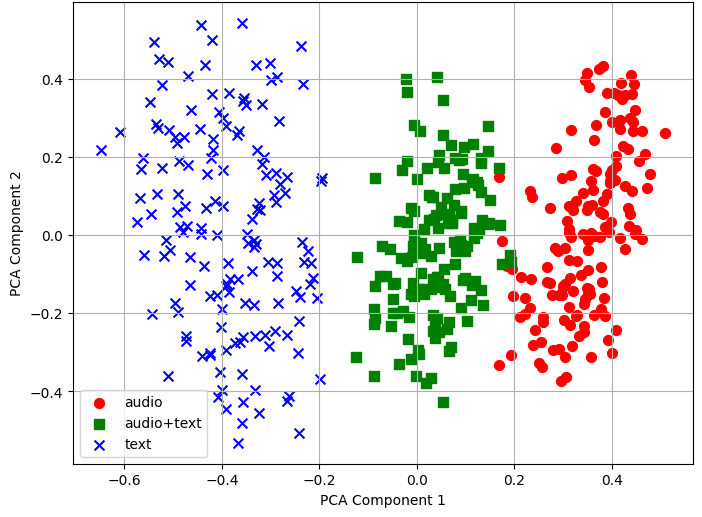} 
         \caption{CLAP Representation}
         \label{fig:sub1}
    \end{subfigure}
    \hfill
    \begin{subfigure}[b]{0.32\textwidth}
         \centering
         \includegraphics[width=\textwidth]{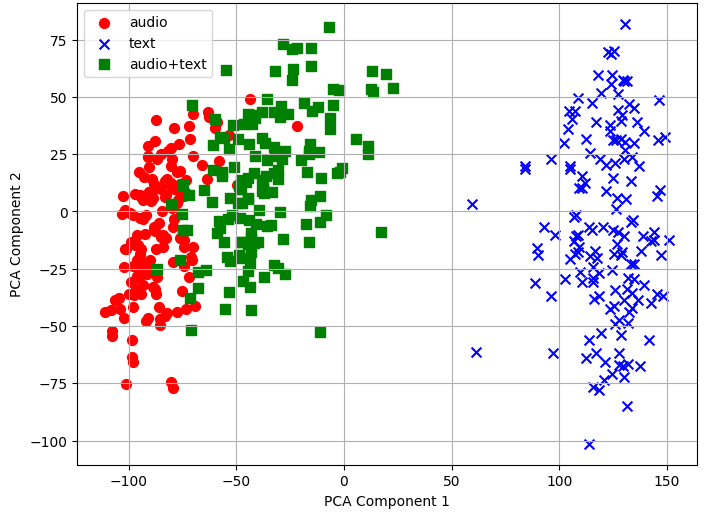} 
         \caption{Original Qwen2audio Representation}
         \label{fig:sub2}
    \end{subfigure}
    \hfill
    \begin{subfigure}[b]{0.32\textwidth}
         \centering
         \includegraphics[width=\textwidth]{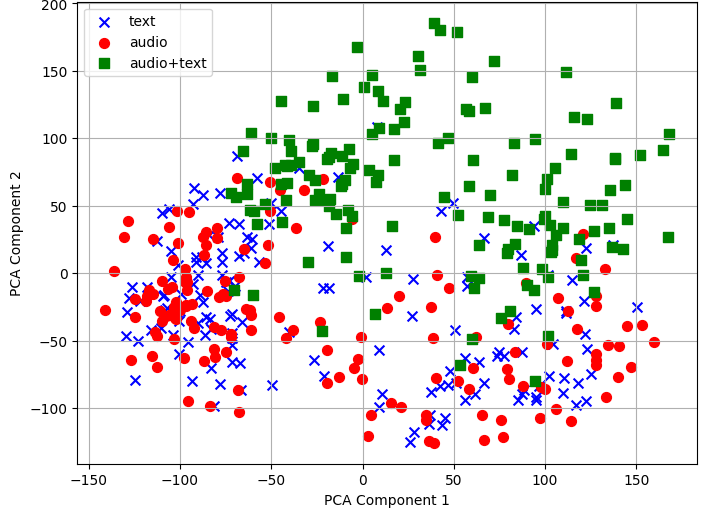} 
         \caption{WavRAG Representation}
         \label{fig:sub3}
    \end{subfigure}
    
    \caption{Comparison of different model representations.}
    \label{fig:representation}
\end{figure*}

To visualize the embedding spaces learned by different models, we randomly selected 150 samples from our test dataset, representing the four retrieval scenarios. Each sample consisted of either an audio clip and its caption, or a group of paired audio-text data. We extracted embeddings for each sample using CLAP, the original Qwen2-Audio model, and our WavRAG retriever. These embeddings were then projected into a two-dimensional space using Principal Component Analysis (PCA).
Figure~\ref{fig:representation} shows the resulting PCA visualizations. In contrast to CLAP and the original Qwen2-Audio, the WavRAG embeddings show no clear separation between modalities. Instead, the audio, text, and combined audio+text embeddings for a given piece of information are closely clustered, indicating that WavRAG consistently represents the same semantic content across modalities.
\begin{table*}[ht]
\centering
\resizebox{\linewidth}{!}{%
  \begin{tabular}{@{}p{3cm}p{12cm}@{}}
    \toprule
    \textbf{Field} & \textbf{Content} \\ \midrule
    id\_a & 18 \\[0.5em]
    caption\_a & Someone is playing a fast melody on a low bansuri flute along with someone playing tablas and a shrutibox in the background. This song may be playing at a live performance. \\[1em]
    id\_b & 3031 \\[0.5em]
    caption\_b & This is an Indian traditional music piece performed live by an orchestra. The orchestra is composed of the bansuri, the sitars and the tabla percussion. There is a vibrant atmosphere in the recording. This piece could be used in the soundtrack of a Bollywood movie. \\[1em]
    rewrite\_knowledge & The bansuri is a side-blown flute originating from the Indian subcontinent, made from a single hollow shaft of bamboo with six or seven finger holes. An ancient musical instrument associated with the pastoral tradition and intimately linked to the love story of Krishna and Radha, it is mentioned in the Vedas and other ancient Hindu texts. The bansuri is revered as Lord Krishna's divine instrument and is often depicted in his iconography. The North Indian bansuri, typically about 14 inches long, is traditionally used as a soprano instrument primarily for accompaniment in lighter compositions including film music. The bass variety (approximately 30 inches, with a tonic typically an octave lower) is preferred in Hindustani classical music. The sound of a bansuri is generated from the resonance of the air column inside it. The length of this column is varied by closing or leaving open a varying number of holes. Half-holes are used to play flat or minor notes. The size of a bansuri affects its pitch. Longer bansuris with a larger bore have a lower range and the shorter ones sound higher. To play the diatonic scale on a bansuri, one needs to find where the notes lie. For example, in a bansuri where Sa or the tonic is always played by closing the first three holes, one can play these notes. The bansuri is typically played in folk and classical music and often accompanies other instruments like the tabla and shrutibox. The tabla is a pair of twin hand drums from the Indian subcontinent. Since the 18th century, the tabla has been the principal percussion instrument in Hindustani classical music, where it may be played solo, as accompaniment with other instruments and vocals, and as a part of larger ensembles. The shrutibox, also known as surpeti, is an instrument that traditionally works on a system of bellows. It is similar to a harmonium and is used to provide a drone in a practice session or concert of Indian classical music. The combination of these instruments is a hallmark of traditional Indian music, often performed live, creating a rich, textured soundscape. \\[1em]
    question\_generation & What are the instruments used, and what is the style of the music being played? \\[1em]
    answer\_generation & The music features a traditional Indian ensemble, highlighted by the distinct sounds of the bansuri, sitar, and tabla. The bansuri, a type of bamboo flute, carries the melody with its characteristic reedy and airy tones, indicative of its use in both folk and classical Indian music. Accompanying the bansuri are the sitar, whose resonant strings add a layer of melodic complexity, and the tabla, providing a rhythmic foundation with its dynamic percussive beats. The interplay of these instruments suggests a live performance setting, capturing the vibrant essence of a traditional Indian music piece. This style is often associated with cultural and cinematic expressions, particularly in Bollywood. \\
    \bottomrule
  \end{tabular}%
}
\caption{Musiccap Sample 1}
\label{musicap1}
\end{table*}

\begin{table*}[ht]
\centering
\resizebox{\linewidth}{!}{%
  \begin{tabular}{@{}p{3cm}p{12cm}@{}}
    \toprule
    \textbf{Field} & \textbf{Content} \\ \midrule
    id\_a & 34 \\[0.5em]
    caption\_a & An acoustic drum is playing a simple rock groove with some drum fills on the toms and snare along with a distorted e-bass and e-guitar playing the same line. Panned to the right side of the speaker you can hear a piano sound playing a little melody. A female voice is singing a happy sounding melody while male backing voices are supporting her. This song may be playing sitting in your room enjoying being at home alone. \\[1em]
    id\_b & 4585 \\[0.5em]
    caption\_b & The low quality recording features a punk song that contains flat male vocal singing over punchy kick and snare hits, shimmering cymbals, wide electric guitars and groovy bass guitar. It sounds energetic, exciting and upbeat---like something you would jump to at concerts. \\[1em]
    rewrite\_knowledge & The audio features a typical rock ensemble, characteristic of the genre's instrumentation and arrangement during its peak popularity in the mid-20th century, particularly from the 1960s through the 1980s. The presence of an acoustic drum set laying down a "simple rock groove" indicates a 4/4 time signature with an emphasis on the backbeat (beats 2 and 4), a foundational element of rock music. The inclusion of drum fills on the toms and snare suggests dynamic variations within the song, a technique used to build tension and release or to transition between sections. The "distorted e-bass and e-guitar playing the same line" points to a common practice in rock music where the bass guitar doubles the guitar riff, creating a heavier, more unified sound. This technique is particularly prevalent in genres like hard rock and heavy metal, which emerged from the broader rock tradition. The distortion effect on the electric instruments is achieved through the use of overdrive or distortion pedals, or by cranking up the amplifier's gain, resulting in a "fuzzy" or "crunchy" tone that is synonymous with rock music. The "piano sound playing a little melody" panned to the right side adds a melodic counterpoint to the rhythm section. The use of panning creates a stereo image, giving the listener a sense of spatial depth. The female voice singing a "happy sounding melody" with male backing vocals suggests a lead and harmony vocal arrangement, a common feature in many rock subgenres.  \\[1em]
    question\_generation & Considering the low quality and punk style of the recording, what is the likely era and subgenre of the rock music being played, and what musical characteristics might be expected in this context? \\[1em]
    answer\_generation & Based on the energetic, exciting, and upbeat nature of the low-quality recording, combined with the punchy drums, shimmering cymbals, wide electric guitars, and groovy bass, it sounds like a classic example of punk rock. This genre often features a raw, stripped-down sound, and the flat male vocals further reinforce this impression. The music's energy and the feeling it evokes, described as something you would "jump to at concerts," are hallmarks of punk's rebellious and lively spirit. Given the characteristics described, the song likely originates from the mid-1970s to early 1980s, a period when punk rock was flourishing and establishing its distinctive sound and ethos. \\
    \bottomrule
  \end{tabular}%
}
\caption{Musiccap Sample 2}
\label{musiccap2}
\end{table*}

\begin{table*}[ht]
\centering
\resizebox{\linewidth}{!}{%
  \begin{tabular}{@{}p{3cm}p{12cm}@{}}
    \toprule
    \textbf{Field} & \textbf{Content} \\ \midrule
    question\_audio & blues/prideandjoy/Pride and Joy (Piano Cover).mp3 \\[0.5em]
    question & what song it is \\[0.5em]
    answer & pride and joy \\[0.5em]
    knowledge\_audio & blues/prideandjoy/Stevie Ray Vaughan \& Double Trouble - Pride and Joy (Official Audio).mp3 \\[0.5em]
    knowledge & "Pride and Joy" is a song by American singer, guitarist and songwriter Stevie Ray Vaughan and his backing band Double Trouble, released in late 1983 by Epic Records. It lists Vaughan as the writer, but actually it is rewritten from a 1962 record called "I Go Into Orbit" by Johnny Acey. The song was released on Stevie's debut studio album Texas Flood (1983). "Pride and Joy" was released as Vaughan's debut single and has become one of his most popular songs. \\
    \bottomrule
  \end{tabular}%
}
\caption{Collected Sample}
\label{collected}
\end{table*}

\begin{table*}[ht]
\centering
\resizebox{\linewidth}{!}{%
  \begin{tabular}{@{}p{3cm}p{12cm}@{}}
    \toprule
    \textbf{Field} & \textbf{Content} \\ \midrule
    id & id10001 \\[0.5em]
    name & A.J.\_Buckley \\[0.5em]
    intro & A.J. Buckley is an Irish-born Canadian actor best known for his role as Ed Zeddmore in the television series \textit{Supernatural} and as Sonny Quinn in the military drama series \textit{SEAL Team}. He has also appeared in numerous other television shows and films, often portraying tech-savvy characters or military personnel. Buckley is recognized for his intense on-screen presence and is also a co-founder of the Paperclip clothing line. He is married with three children. \\[1em]
    audio\_samples & J9lHsKG98U8/00016.wav; Y8hIVOBuels/00007.wav \\[1em]
    question & Based on the voice, who do you think this is? \\[1em]
    answer & A.J.\_Buckley \\
    \bottomrule
  \end{tabular}%
}
\caption{Voxceleb Sample}
\label{vox}
\end{table*}

\begin{table*}[ht]
\centering
\resizebox{\linewidth}{!}{%
  \begin{tabular}{@{}p{3cm}p{12cm}@{}}
    \toprule
    \textbf{Field} & \textbf{Content} \\ \midrule
    id\_a & 7 \\[0.5em]
    filename\_a & 002\_78\_rpm\_vinyl\_noise\_44\_16\_lossless.wav \\[0.5em]
    captions\_a & \begin{tabular}[t]{@{}l@{}}
      1. Static on a stereo or a similar device.\\[0.2em]
      2. Consistent, uninterrupted static with the occasional pop like a speaker going out.\\[0.2em]
      3. White noise static is droning and crackling in the background.\\[0.2em]
      4. With the occasional pop like a speaker going out consistent uninterrupted static sounds.\\[0.2em]
      5. White noise static drones and crackles in the background.
    \end{tabular} \\[1em]
    id\_b & 2310 \\[0.5em]
    filename\_b & Vinyl record noise.wav \\[0.5em]
    captions\_b & \begin{tabular}[t]{@{}l@{}}
      1. A record has reached the end of a song and is waiting to start over.\\[0.2em]
      2. A turntable turns with no needle touching record.\\[0.2em]
      3. An electronic hiss and pop is all the speaker emits.\\[0.2em]
      4. The cracking of a small fire very close to the receiver.\\[0.2em]
      5. The crackling of a small fire very close to the receiver.
    \end{tabular} \\[1em]
    rewrite\_knowledge & Static in audio refers to a type of noise characterized by a random hissing or crackling sound, often perceived as white noise or a similar broadband noise. In the context of audio equipment like stereos, static can originate from various sources. It can be caused by atmospheric interference affecting radio wave reception, particularly in older analog radio systems, or electronic interference within the device itself, potentially due to faulty components or poor shielding. Stereo systems, short for stereophonic sound systems, are designed to create an illusion of multi-directional sound perspective, enhancing the listening experience by distributing audio through two or more channels. Historically, stereos became popular for home listening in the mid-20th century, often incorporating various input sources such as radio tuners, phonograph record players, and tape players. When static is heard through a stereo system, especially from the speaker output rather than an external source like a radio broadcast, it often indicates a problem within the amplifier or the audio signal processing circuitry of the stereo itself. This kind of internally generated static is distinct from the intentional use of static or noise in sound design or electronic music for artistic effect. \\[1em]
    question\_generation & What sound is it? \\[1em]
    answer\_generation & The sound you're hearing likely originates from a mechanical audio playback device, possibly one utilizing a physical medium for sound storage. The pause and the nature of the sound suggest the equipment has reached the end of a discrete segment of audio, such as a track on a record. Older audio technologies often employed physical mechanisms to play and transition between recordings. The sound is characteristic of a system waiting for a manual or automatic command to initiate the next segment of audio playback, a process that is distinct from the seamless transitions found in modern digital audio systems. This pause, accompanied by mechanical sounds, is a hallmark of certain analog audio formats where playback is not continuous but segmented by the physical structure of the recording medium. \\
    \bottomrule
  \end{tabular}%
}
\caption{Clotho Sample}
\label{clotho}
\end{table*}

\end{document}